\documentclass[aps,pra,superscriptaddress,showpacs]{revtex4}
\usepackage{graphicx}

\newcommand{\Si}{\mathop{\rm Si}\nolimits}
\newcommand{\Ci}{\mathop{\rm Ci}\nolimits}

\begin{document}

\title{Fluctuations of the number of particles within a given volume in cold quantum gases}

\author{G. E. Astrakharchik}
\affiliation{Departament de F\'{\i}sica i Enginyeria Nuclear, Campus Nord B4-B5, Universitat Polit\`ecnica de Catalunya, E-08034 Barcelona, Spain}

\author{R. Combescot}
\affiliation{Laboratoire de Physique Statistique, Ecole Normale Sup\'erieure, 24 rue Lhomond, 75231 Paris Cedex 05, France}

\author{L. P. Pitaevskii}
\affiliation{Dipartimento di Fisica, Universit\`a di Trento and BEC-INFM, I-38050 Povo, Italy; Kapitza Institute for Physical Problems, 119334 Moscow, Russia}

\date{\today}

\begin{abstract}
In ultracold gases many experiments use atom imaging as a basic observable. The resulting image is averaged over a number of realizations and mostly only this average is used. Only recently the noise has been measured to extract physical information. In the present paper we investigate the quantum noise arising in these gases at zero temperature. We restrict ourselves to the homogeneous situation and study the fluctuations in particle number found within a given volume in the gas, and more specifically inside a sphere of radius $R$. We show that zero-temperature fluctuations are not extensive and the leading term scales with sphere radius $R$ as $R^2\ln R$ (or $\ln R$) in three- (or one-) dimensional systems. We calculate systematically the next term beyond this leading order. We consider first the generic case of a compressible superfluid. Then we investigate the whole Bose-Einstein-condensation (BEC)--BCS crossover crossover, and in particular the limiting cases of the weakly interacting Bose gas and of the free Fermi gas.
\end{abstract}


\maketitle

\section{Introduction}

In recent years major progress has been achieved in realizations of highly controlled low temperature quantum gases. One of the basic techniques, which permits the investigation of the properties of cold atomic clouds, is imaging of the condensate on a charge-coupled device camera. Usually the noise, present in each image of a particular realization of the cloud, is ignored and the overall density is fitted by a smooth function. This enables the extraction of the density profile and size of the cloud. Only recently a new generation of experiments has appeared \cite{Greiner05} in which the noise itself is analyzed. It was shown \cite{Greiner05} that the study of the correlation functions in real space reveals correlations due to the presence of fermionic dimers in the BEC-BCS crossover problem. Naturally, proper control of these kinds of experiments requires a good theoretical understanding of the noise in these systems, and more specifically of the influence of quantum properties on the noise since these properties are the major interest of these systems. In the present paper we point out that these correlations at low temperatures indeed show interesting quantum properties and permit the extraction of additional information on the interactions in the system, for example, at unitarity. Since the systems of interest display the BEC-BCS crossover, we investigate in detail density fluctuations throughout this crossover.

The problem of fluctuations was addressed in a number of theoretical studies \cite{Price54,Giorgini98,Belzig05,Castin06}. Recently Belzig {\it et al.} \cite{Belzig05} studied fluctuations $\left\langle \delta N^2\right\rangle =\left\langle (N-\bar N)^2\right\rangle$ in the BEC-BCS crossover at zero temperature using BCS theory. In their approach number fluctuations in a large enough volume grow linearly with the volume. It was predicted that, deep in the BCS regime, fluctuations are proportional to the value of the gap and are vanishingly small. However, because of their use of the simple BCS wave function, they did not take into account the collective mode, which as we will see plays a crucial role. Number fluctuations of a Bose gas in a sphere were studied with logarithmic accuracy in \cite{Giorgini98}. Classical thermodynamic fluctuations are known to be extensive, {\it i.e.} additive with respect to the volume. On the contrary, at zero temperature fluctuations grow less rapidly, which is physically reasonable since the thermal source of fluctuations is no longer present. On the other hand there are still quantum fluctuations. It was found that these fluctuations at zero temperature are not additive and grow less rapidly than the volume. In the present paper we extend this work in two directions. First we go to the next significant order in the dependence of fluctuations on volume by performing the calculation with nonlogarithmic accuracy. Next we extend our study to the BEC-BCS crossover, which allows us in particular to cover at the same time the case of free bosons and free fermions. Fluctuations in systems with reduced dimensionality are analyzed by studying properties of a one-dimensional free Fermi gas.

The paper is organized as follows. In Sec.~\ref{gen} we recall as an introduction to our calculations the basics of the fluctuation calculations. Then, in Sec.~\ref{secSound}, we address briefly the case of a general compressible superfluid and recall that the dominant term is directly linked to long wavelength excitations, i.e. phonons. After this, in Sec.~\ref{secBosons} we proceed to a more precise calculation in the case of weakly interacting bosons and obtain explicitly the next term beyond the dominant logarithmic one. Section~\ref{secBCSBEC} is devoted to the extension of this calculation to the BEC-BCS crossover, making use of the dynamical BCS model. Finally the case of the ideal Fermi gas is handled exactly in section~\ref{secIFG}, both for a three-dimensional and a one-dimensional space. In the latter case we also comment on non zero temperature properties.

Notice, that we will consider Bose systems only in the presence of interparticle interactions. Fluctuations in an ideal BEC gas are quite specific, due to the infinite compressibility of the system. A reader can find a comprehensive review of this subject in \cite{Ziff77}.

\section{Fluctuations in a sphere\label{gen}}

In a homogeneous system the average number of particles $\bar N$ in a given volume within the system is merely fixed by the density and the size of the volume.
In contrast, the particle number fluctuations depend on interaction properties of the gas. The strength of number fluctuations is quantitatively described by the dispersion $\left\langle \delta N^2\right\rangle =\left\langle (N-\bar N)^2\right\rangle$. These fluctuations are closely related to the density-density static correlation function $\bar n\nu(|\mathbf{r}_1\mathbf{-r}_2|) \equiv \langle\delta n(\mathbf{r}_1)\delta n(\mathbf{r}_2)\rangle$:
\begin{eqnarray}
\left\langle \delta N^2\right\rangle = \bar n\int\limits_{V}\int\limits_{V}\nu(|\mathbf{r}_1\mathbf{-r}_2|)\;d\mathbf{r}_1d\mathbf{r}_2, \label{dN}
\end{eqnarray}
where $\bar n = \langle n(r)\rangle$ denotes the average density in a homogeneous system, $\delta n(\mathbf{r})=n(\mathbf{r})-\bar n$ and $V$ is the volume in which the fluctuations are measured.

In homogeneous systems (to which we will restrict ourselves in the following) isotropy makes it natural to investigate number fluctuations in the situation where the volume $V$ is a sphere of radius $R$. Then the mean number of particles is $\bar N = \frac{4}{3}\pi R^3\bar n$. The double integration in Eq.~(\ref{dN}) can be partially performed resulting in a one-dimensional integral:
\begin{eqnarray}
\left\langle \delta N^2\right\rangle=4\pi {\bar n}\int\limits_0^{2R}\nu(r)\tau(r)r^2\;dr, \label{fluct}
\end{eqnarray}
where we have introduced the overlapping volume $\tau(r)= \int_{V} \int_{V} d\mathbf{r}_1d\mathbf{r}_2 \delta(\mathbf{r}_1-\mathbf{r}_2-\mathbf{r})$ of two spheres of radius $R$, with centers separated by $r$, which is given explicitly by
\begin{eqnarray}
\tau(r)=\frac{4\pi R^3}{3}-\pi rR^2+\frac{\pi r^3}{12}. \label{tau}
\end{eqnarray}

The density-density static correlation function (pair distribution function) is related to the probability of finding two particles separated by a distance $r$. The function $\nu(r)$ is defined in coordinate space. Its counterpart in momentum space is the static structure factor $S(k)$, which we define precisely as
\begin{eqnarray}
\nu(\mathbf{r})=\int e^{i\mathbf{kr}}S(k)\frac{d^3k}{(2\pi)^3}. \label{nu:Sk}
\end{eqnarray}

Below we will use directly our knowledge of $S(k)$ to calculate $\left\langle \delta N^2\right\rangle$. It is useful to introduce also the Fourier transform of $\tau(r)$ which is immediately obtained from its above expression as $T(k)= \int_{V} \int_{V} d\mathbf{r}_1d\mathbf{r}_2e^{i\mathbf{k}(\mathbf{r}_1-\mathbf{r}_2)}$ and is given by
\begin{eqnarray}\label{TFnu}
T(k)=\frac{16 \pi ^2}{k^6}\left[\sin(kR)-kR \cos(kR)\right]^2.
\end{eqnarray}
Then Eq.~(\ref{fluct}) is replaced by
\begin{eqnarray}\label{dNTF}
\left\langle\delta N^2\right\rangle = \frac{{\bar n}}{2\pi ^2}\int_{0}^{\infty} S(k) T(k)k^2\;dk.
\end{eqnarray}

If the $f$-sum rule is exhausted by the collective mode branch of the excitations, the static structure factor can be found from the Feynman relation
\begin{eqnarray}
S(k) = \frac{\hbar k^2}{2m\omega_s(k)}, \label{Feynman}
\end{eqnarray}
where $m$ is the mass of an atom and $\omega_s(k)$ is the excitation energy.

The knowledge of the single branch excitation spectrum in a weakly interacting Bose gas allows us to calculate, using Eqs.~(\ref{dN})-(\ref{Feynman}), the dependence of the density fluctuations $\left\langle \delta N^2\right\rangle$ on the strength of interactions and the radius of the sphere.

\section{Compressible superfluid\label{secSound}}

The long wavelength behavior of the static structure factor in any compressible superfluid is universal and corresponds to the presence of sound waves with velocity $c$ and linear excitation spectrum $\omega_{ph}(k)=k c$. By using the Feynman relation, Eq.~(\ref{Feynman}), which in the general case is valid in the $k \to 0$ limit, we immediately find a linear dependence of the static structure factor $S(k) = \hbar k/(2mc)$. Substitution of $S(k)$ into Eq.~(\ref{nu:Sk}) gives the long range decay of the density-density correlation function $\nu(r)\propto 1/r^4$ (see Sec.~87 in \cite{Lifshitz80}). There are three terms in Eq.~(\ref{tau}). The contribution of the constant term can be calculated using the identity
 $4\pi\int_0^{\infty}\nu(r)r^2dr =S(k)\left|_{k=0}\right. = 0$.
Thus this contribution scales as $R^3\int_0^{2R}\nu(r)r^2dr\propto -R^3\int_{2R}^\infty\frac{dr}{r^2}\propto R^2$. The linear term causes a logarithmic ultraviolet divergence for $r\rightarrow0$ in the integral of Eq.~(\ref{fluct}). This divergence is due to a failure of the phononic description at large $k$ and is avoided \cite{Giorgini98} by truncation of the integral at some distance $\xi$ of the order of the healing length. This results in a logarithmic dependence of $\left\langle\delta N^2\right\rangle$ on $R$, since $R^2\int_0^{2R}\nu(r)r^3\;dr\propto R^2\int_\xi^{2R}\frac{dr}{r}=R^2\ln(R/\xi)$. The last term is already converging and scales as $R^2$ since $R\int_0^R\nu(r)r^5dr\propto\int_0^Rr\;dr\propto R^2$. Hence the leading contribution to the integral comes from the term in $\tau(r)$ linear in $r$. The final expression reads as \cite{Giorgini98}
\begin{eqnarray}
\left\langle\delta N^2\right\rangle =\frac{2{\bar n}\hbar}{mc}R^2\ln\left({\mathcal C}\; \frac{R}{\xi}\right),\qquad R\gg\xi \label{universal}
\end{eqnarray}
with the coefficient ${\mathcal C}$ inside the logarithm undetermined, as it depends on the unspecified short-range behavior.

The result Eq.~(\ref{universal}) does not depend on statistics and is valid both in fermionic and bosonic superfluids. Thus at $T=0$ the increase of $\langle\delta N^2\rangle$ when $R\to\infty$ follows the law $R^2\ln R$. In contrast, at finite temperature according to the classical thermodynamic equation, $\left\langle \delta N^2\right\rangle = -k_B T{\bar n}(\partial V/\partial P)_T \propto R^3$, which is faster compared to Eq.~(\ref{universal}). This was first noted in \cite{Giorgini98}.

Notice that at finite, but low temperature Eq.~(\ref{universal}) is valid at the distances $R\ll \hbar c/(k_BT)$. For distances satisfying the opposite inequality, the density-density correlation function decreases exponentially (see \cite{Lifshitz80}, Problem~1 to Sec.~87) and the classical result for number fluctuations is valid.

It is important to note that we consider the fluctuations of the {\em total} number of atoms in a given volume. Fluctuations in the number of atoms in the condensate and in the number of atoms out of condensate separately increase faster with $R$: $\left\langle \delta N_0^2\right\rangle \propto R^3$ at $T=0$ and $\propto R^4$ at finite $T$ (for detailed discussions see \cite{Giorgini98} and \cite{Kocharovsky00}).

\section{System of weakly interacting Bosons\label{secBosons}}

One can easily calculate the coefficient ${\mathcal C}$ in a system of weakly interacting bosons with mass $m_B$, where the average boson density ${\bar n}_B$ satisfies ${\bar n}_Ba^3\ll 1$ ($a$ being $s$-wave scattering length). These conditions correspond to the present experimental situation for trapped BEC gases, which are perfectly described by the Bogoliubov theory. The excitation spectrum in this theory has the form $\omega_B(k) = \sqrt{(kc)^2+(\frac{\hbar k^2}{2m_B})^2}$ and sound velocity is $c=\sqrt{4\pi {\bar n}_Ba^3}\;\hbar/m_Ba$. The static structure factor is then found from the Feynman relation Eq.~(\ref{Feynman}) which is exact in the framework of the Bogoliubov approximation
\begin{eqnarray}
S_B(k) = \frac{\frac{\hbar^2k^2}{2m_B}} {\sqrt{(\hbar k c)^2+(\frac{\hbar^2k^2}{2m_B})^2}}. \label{Sk:Bogoliubov}
\end{eqnarray}
The small-momentum part $k\rightarrow0$ describes phonons $S_B(k)\simeq \hbar k/(2m_Bc)$, while the large-momentum part $k \to \infty$ corresponds to free particles with energy given by $E_k = \hbar^2k^2/2m_B$ and static structure factor $S_B(k)\to 1$.

By substituting Eq.~(\ref{Sk:Bogoliubov}) into Eqs.~(\ref{fluct})-(\ref{nu:Sk}) we obtain the density fluctuations in a weakly interacting Bose gas (details of the calculation are given in Appendix~\ref{mod}):
\begin{eqnarray}
\left\langle\delta N^2\right\rangle =\frac{2 {\bar n}_B\hbar}{m_Bc}R^2\ln\left(8e^{\gamma-1}\frac{Rm_Bc}{\hbar}\right), \label{BG}
\end{eqnarray}
where $\gamma=0.577...$ is Euler's constant. We see explicitly that knowledge of the short wavelength physics is important for the calculation of the coefficient inside the logarithm. In our case this is done by assuming the Bogoliubov excitation spectrum.

Notice that our theory is valid for a uniform gas. However, it can be applied to a small volume inside a trapped BEC gas in the Thomas-Fermi regime under the condition $R \ll R_0$, where $R_0$ is the condensate radius.

\section{\textbf{Superfluid } system of two-component fermions\label{secBCSBEC}}

We now consider the calculation for a system with two fermionic species in the BEC-BCS crossover within the dynamical BCS model. As we have mentioned it is convenient to use the numerical results which have already been obtained for $S(k)$ in the study of molecular signatures in this crossover \cite{Combescot06}. Since we now deal with two kinds of fermions with the same mass $m_F$ and equal populations, say spin-up and spin-down fermions, we first have to take this into account in the general formulation of our problem outlined in Sec.~\ref{gen}. Following the same steps it is easy to see that the total number fluctuation is still essentially given by Eq.~(\ref{dNTF}):
\begin{eqnarray}\label{dNTF1}
\left\langle \delta N^2\right\rangle = \frac{{\bar n}_{\mathrm{tot}}}{2\pi^2}\int\limits_{0}^{\infty} S(k) T(k)k^2\;dk
\end{eqnarray}
provided we use the total density ${\bar n}_{\mathrm{tot}} =2{\bar n}$, where we keep the notation ${\bar n} \equiv {\bar n}_{\uparrow}={\bar n}_{\downarrow}$ for the single species average density. Here $S(k)$ is the static structure factor defined by $S(k)=S_{\uparrow \uparrow}(k)+S_{\uparrow \downarrow}(k)$, which has been calculated in Ref.~\cite{Combescot06} and is again related to the sound velocity $c$ in the limit $k \rightarrow 0$ by $S(k) = \hbar k/(2m_F c)$. Naturally we now have $\delta N=\delta N_{\uparrow} + \delta N_{\downarrow}$.

\begin{figure}[tbp]
\begin{center}
\includegraphics[angle=-90,width=0.5\columnwidth]{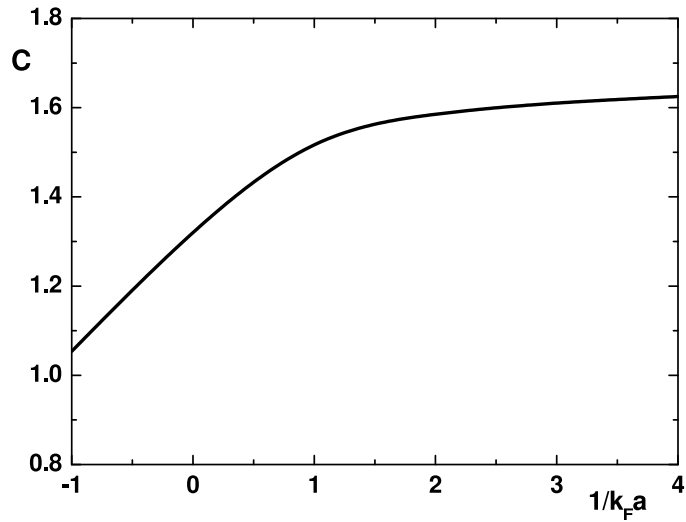}
\end{center}
\caption{Two-component Fermi gas in the BCS-BEC crossover, where the coefficient $C$ in the logarithm is defined as $\langle\delta N^2\rangle = 8 {\bar n} \alpha R^2 [\ln(R/\alpha )+C]$ with $\alpha = \hbar /(2 m_F c)$.} \label{FigC}
\end{figure}

\begin{figure}[tbp]
\begin{center}
\includegraphics[angle=-90,width=0.5\columnwidth]{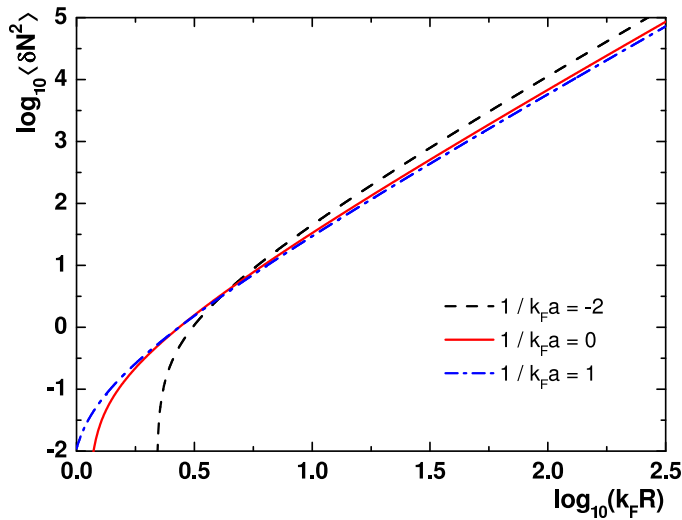}
\end{center}
\caption{(Color online) Density fluctuations in two-component Fermi gas in the BCS-BEC crossover: dashed line, $1/k_Fa=-2$ (BCS regime); solid line, $1/k_Fa=0$ (unitary regime); and dash-dotted line, $1/k_Fa=1$ (BEC regime). } \label{FigBCSBEC}
\end{figure}

Although $S(k)$ is known numerically, brute force numerical calculation is not a good choice and it is better to extract analytically the $R^2 \ln R$ behavior which comes from the low $k$ domain, as we have seen quite generally in Sec.~\ref{secSound}. This is most conveniently achieved by introducing a simple model $S_0(k)$ for the structure factor which can be handled analytically. Then the difference with the actual $S(k)$ is handled numerically. We take for $S_0(k)$ a model which behaves as $S(k)$ in the two limits $k\rightarrow0$ and $k\rightarrow \infty$, in order that the difference $S(k)-S_0(k)$ goes to zero in these two limits. This simplifies both the general analysis and the numerical calculation. We choose:
\begin{eqnarray}\label{model}
\left\{\begin{array}{cc}\hspace{2mm} S_0(k)=\alpha k, & \hspace{15mm} k < \frac{1}{\alpha}
\\ & \\S_0(k)=1, & \hspace{15mm} k > \frac{1}{\alpha }\end{array}\right.,
\end{eqnarray}
where we take the coefficient $\alpha = \hbar /(2m_F c)$ in order to recover the proper behavior for $k\rightarrow0$. The details of the analytical calculation of $\left\langle \delta N^2\right\rangle$ for this model are given in Appendix~\ref{mod}. The result is
\begin{eqnarray}\label{resmod}
\left\langle \delta N^2\right\rangle_{\mathrm {mod}} =8{\bar n}\alpha R^2\ln \frac{2 e^\gamma R}{\alpha }=8{\bar n}\alpha R^2 \left[\;\ln \frac{R}{\alpha } + 1.270...\;\right].
\end{eqnarray}
Then we are left with the calculation of
\begin{eqnarray}
\left\langle\delta N^2\right\rangle-\left\langle \delta N^2\right\rangle_{\mathrm {mod}}= 16{\bar n}\!\left[\int\limits_{0}^{1/\alpha}\!\!dk\frac{S(k)-\alpha k}{k^4} \left[\sin(kR)-kR \cos(kR)\right]^2\right. \left. \!+\!\int\limits_{1/\alpha}^{\infty}\!\!dk\frac{S(k)-1}{k^4} \left[\sin(kR)-kR \cos(kR)\right]^2\right].
 \end{eqnarray}
In both terms the dominant contribution in the limit $R \rightarrow \infty$ comes from the $kR \cos(kR)$ term, and moreover, writing $\cos^2(kR)=(1/2)(1+\cos(2kR))$, we see that only the $1/2$ is left at dominant order. In other words we may replace $\cos^2(kR)$ by its average $1/2$. This leads to
\begin{eqnarray}
\hspace{-30mm}\left\langle \delta N^2\right\rangle-\left\langle \delta N^2\right\rangle_{\mathrm {mod}}= 8{\bar n} R^2 \left[\;\int\limits_{0}^{1/\alpha} dk\frac{S(k)-\alpha k}{k^2} + \int\limits_{1/\alpha }^{\infty} dk\frac{S(k)-1}{k^2} \; \right].
\end{eqnarray}
We see in particular that the first integration no longer gives rise to a divergence in the limit $k \rightarrow 0$, which is present when one has $S(k)$ instead of $S(k)-\alpha k$ in the numerator and is responsible for the $R^2 \ln R$ contribution. Hence the integrations are regular in both the limits $k \rightarrow 0$ and $k \rightarrow \infty$ and are easily performed numerically once the numerical values of $S(k)$ are known. The result is presented in Fig.~\ref{FigC}. We have written the number fluctuation in the general form $\langle\delta N^2\rangle = 8{\bar n} \alpha R^2 \left[\ln(R/\alpha )+C\right]$, where $\alpha = \hbar /(2 m_F c)$ and we have displayed the coefficient $C$ as a function of $1/(k_Fa)$. We have also plotted in Fig.~\ref{FigBCSBEC}, from this asymptotic result, the behavior of the number fluctuation as a function of the radius of the sphere $R$, for various values of $1/(k_Fa)$. Our extension of this figure toward fairly low values of $k_FR$ is justified by our finding in the case of the free Fermi gas, considered in the next section.

It is naturally of interest to consider the BEC limit $1/(k_Fa) \rightarrow +\infty$ and to compare the numerical result with the analytical one found in Sec.~\ref{secBosons}. First the number $\delta N$ considered in Sec.~\ref{secBosons} corresponds to fluctuations of bosons. Since in the BEC limit the bosons we end up with contain two fermions, we have to multiply the result of Sec.~\ref{secBosons} by four in order to obtain a fermion fluctuation which can be compared with Eq.~(\ref{BG}). Hence we expect to find in the BEC limit
\begin{eqnarray}\label{bosf}
\left\langle\delta N^2\right\rangle =\frac{4 {\bar n} \hbar}{m_Fc}R^2\ln\left(16e^{\gamma-1}\frac{Rm_Fc}{\hbar}\right)
\end{eqnarray}
since with our convention we have ${\bar n}_B = {\bar n}$, while $m_B=2m_F$. Taking into account $\alpha =\hbar/(2m_Fc)$ we see that the coefficients in front of $R^2 \ln R$ agree as they should. Next from Eq.~(\ref{bosf}) we should find $C=3 \ln 2 +\gamma-1=1.657...$ for the coefficient $C$ defined above, while we see from Fig.~\ref{FigBCSBEC} that for $1/k_Fa=4$ the result is $C=1.625...$. Since we are not yet in the strong BEC limit $1/k_Fa \rightarrow \infty$, the agreement is excellent. It is worth noting that a very good agreement with the BEC limit is already found starting from $1/k_Fa \simeq 1$.

\begin{figure}[tbp]
\begin{center}
\includegraphics[angle=0,width=0.4\columnwidth]{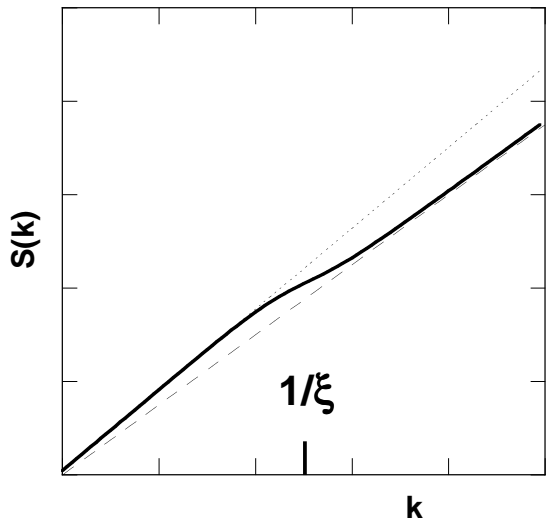}
\end{center}
\caption{Qualitative behavior of $S(k)$ at low $k$ in the limit of large and negative $1/k_Fa$. In the limit $k \rightarrow 0$, $S(k)=\sqrt{3}k/2k_F$ (dotted line). For $k \approx 1/\xi$, where $\xi$ is the coherence length, this behavior switches to the free Fermi gas result $S(k)=3k/4k_F$ (dashed line).} \label{xover}
\end{figure}

It would also seem natural to investigate in the same spirit the BCS limit $1/k_Fa \rightarrow -\infty$. However, it is well-known that this limit is somewhat singular. Indeed, when $1/k_Fa$ goes toward $-\infty$, $S(k)$ becomes identical to the structure factor of the free Fermi gas; but in the small $k$ range this occurs in a singular way. For a free Fermi gas, $S(k) \simeq 3 k/4k_F$ for $k \rightarrow 0$, that is $\alpha = 3/4k_F$, but the sound velocity we will find for large negative $1/k_Fa$ is given by $c=\sqrt{(n/m_F)\partial \mu /\partial n}=\hbar k_F / m_F \sqrt{3}$ which means that the relation $\alpha = \hbar / 2m_Fc$ does not apply. Hence, we can not recover the free Fermi gas in the limit $1/k_Fa \rightarrow -\infty$ since we will already have a disagreement at the level of the coefficient of the $R^2 \ln R$ term. The basic physical reason is that the above expression for the (first) sound velocity implies naturally thermodynamic equilibrium, while in the non-interacting gas the relaxation time is infinite which prevents the existence of first sound. On the other hand it is known that in the superfluid state this singular behavior disappears and first sound exists with a velocity related to compressibility and given by the above general formula. For large (but finite) negative $1/k_Fa$, $S(k)$ has the following anomalous behavior, which is displayed qualitatively in Fig.~\ref{xover}. For $k \rightarrow 0$ one has indeed $S(k) \simeq \hbar k/2m_Fc=\sqrt{3}k/2k_F$. However this behavior holds only in the small range $q \lesssim 1/\xi \approx \Delta m_F/\hbar k_F$, where $\xi$ is the coherence length and $\Delta $ the BCS gap. Beyond this range one recovers the free Fermi behavior $S(k) \simeq 3 k/4k_F$; but, since $\Delta \rightarrow 0$ when $1/k_Fa \rightarrow -\infty$, all this structure for $S(k)$ occurs in a range of vanishing size in this limit. If we insist on extending our above calculations in this range, the change of slope of $S(k)$ at small $k$ we have just described will produce a divergent result for our coefficient $C$ in the limit $1/k_Fa \rightarrow -\infty$. Another way to state this result is to say that the two limits $1/k_Fa \rightarrow -\infty$ and $R \rightarrow \infty$ do not commute. One finds the free Fermi gas limit if one keeps $R$ finite and first let $1/k_Fa \rightarrow -\infty$, but not when these limits are exchanged, as is clear by considering the coefficient of the $R^2 \ln R$ term.

After having made this point we calculate $\left\langle\delta N^2\right\rangle$ nevertheless for completeness in the following section for the free Fermi gas.

\section{Ideal Fermi gas\label{secIFG}}

\begin{figure}[tbp]
\begin{center}
\includegraphics[angle=-90,width=0.5\columnwidth]{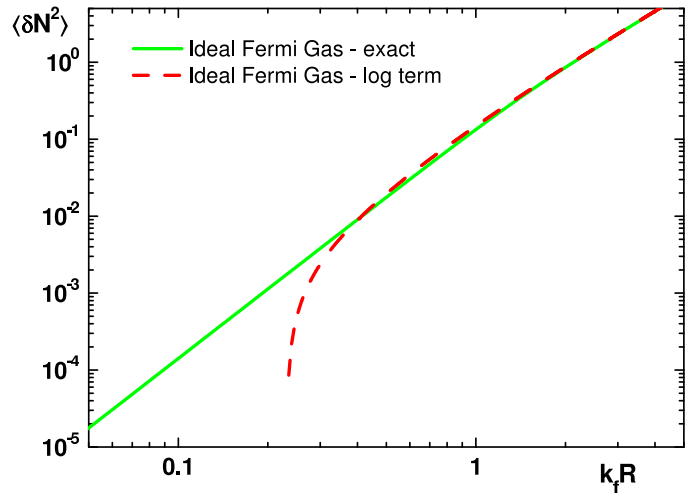}
\end{center}
\caption{(Color online) Number fluctuations in an ideal Fermi gas. Solid green line - exact evaluation of the integral Eq.~(\protect\ref{IFGexact}) and dashed red line - leading logarithmic contribution Eq.~(\protect\ref{IFGlog})} \label{FigIFG}
\end{figure}

In the case of ideal Fermi gas it is possible to perform a fully explicit calculation of $\delta N^2$. We can then compare the logarithmic-accuracy expression with this exact result. In an ideal Fermi gas there are no correlations between atoms of different spin orientations, and hence ${\bar n} \nu_{\uparrow \downarrow}(r) \equiv \langle (n_{\uparrow}({\bf r})-{\bar n}) (n_{\downarrow}({\bf 0})-{\bar n})\rangle=0$ and $\left\langle\delta N^2_{\uparrow\downarrow}\right\rangle \equiv \left\langle (N_{\uparrow} - {\bar N}_{\uparrow}) (N_{\downarrow} - {\bar N}_{\downarrow})\right\rangle=0$. The same-spin density-density correlation function of an ideal Fermi gas at zero temperature is known \cite{Pines89} explicitly and manifests decaying oscillations \cite{rem} with a period defined by the Fermi momentum $k_F = (6\pi^2 {\bar n})^{1/3}$:
\begin{eqnarray}
\nu_{\uparrow\uparrow}(r)=\delta^3({\bf r}) -\frac{9{\bar n}}{(k_{F}r)^4}\left(\frac{\sin k_{F}r}{k_{F}r}-\cos k_{F}r\right)^2, \label{nu:IFG}
\end{eqnarray}
describing the Fermi hole due to Pauli exclusion, with the free particle behavior being recovered in the $r\rightarrow0$ range. Then density fluctuations inside a sphere are given by Eq.~(\ref{fluct}). The integration can be performed exactly in terms of special functions and the full expression is given in Appendix~\ref{idealferm} by formula (\ref{IFGexact}). Here we are interested in the long-range asymptotic. The leading term is
\begin{eqnarray}
\left\langle \delta N_{\uparrow \uparrow }^2\right\rangle =\frac{(k_{F}R)^2}{2\pi^2}\ln \left[4e^{\gamma-\frac{1}2}k_{F}R\right]+\mathcal{O}(k_{F}R). \label{IFGlog}
\end{eqnarray}
The comparison of the asymptotic behavior Eq.~(\ref{IFGlog}) with the exact result (\ref{IFGexact}) is shown in Fig.~\ref{FigIFG}. An important observation is that this long-range asymptotic works very well for distances as small as $k_{F}R~\approx 1$. The coefficient under the logarithm is $4e^{\gamma -\frac{1}{2}} \simeq 4.32$.

Of course, the general logarithmic accuracy expression Eq.~(\ref{universal}) coincides with the direct calculation Eq.~(\ref{IFGlog}) once expressed in appropriate units. We note that for an ideal Fermi gas one should use $c = 2c_F/3m$ (where $c_F=\hbar k_F/m$ is the Fermi velocity) as the expression for the speed of sound, as this quantity enters in the long-range asymptotics of the density-density correlation function and corresponds to the slope of the linear behavior for small momenta in the static structure factor. This velocity is different from the speed calculated through the compressibility $mc^2 = n \partial\mu/\partial n$, which leads instead to $c = c_F/\sqrt{3}$. The difference is quite natural because there are no sound-like excitations in an ideal Fermi gas. This difference is absent in bosonic systems and the system of ideal low-dimensional fermions.

At unitarity the only relevant length scale is again fixed by the Fermi momentum (or density), while the chemical potential is obtained from the non interacting one by multiplication by the coefficient $\xi \simeq 0.42(1)$ \cite{MC}. Accordingly the speed of sound is given by $c=\sqrt{\xi}\hbar k_{F}/\sqrt{3}m$. We note that the value of $\xi$ can be extracted from experiment by a precise measurement of the total fluctuations, since at unitarity we expect
\begin{eqnarray}
\left\langle\delta N^2\right\rangle =\frac{(k_{F}R)^2}{\sqrt{\xi}\pi^2}\ln\left[{\mathcal C}k_{F}R\right].
\end{eqnarray}

Analogous results can be obtained for an ideal one-dimensional Fermi gas in a box of size $2R$, as considered in detail in Appendix~\ref{onedim}, where the nonzero temperature case is also considered. The fluctuations are given by
\begin{eqnarray}
\left\langle\delta N^2\right\rangle=2\bar{n}\int\limits_0^{2R}\nu(r)\tau(r)\;dr, \label{fluct1D}
\end{eqnarray}
where $\tau(r)=2R-r$ and $\bar{n}$ is the particle density by unit length. For an ideal Fermi gas the density-density correlation function at zero temperature is $\nu(r)=\delta(r)-\bar{n} \sin^2(k_F r)/(k_F r)^2$, where $k_F = \pi \bar{n}$ is the one-dimensional Fermi momentum. The leading term is logarithmic and, with inclusion of the next order correction, the result is found to be:
\begin{eqnarray}
\left\langle \delta N^2\right\rangle^{IFG}_{1D}=\frac{1}{\pi^2}\ln(4e^{\gamma+1}k_F R).
\end{eqnarray}
The complete expression for $\left\langle \delta N^2\right\rangle$ is given by formula (\ref{dN2_1D_T0}).

The low temperature expression for the density-density correlation function $\nu(r)$ is obtained in Appendix~\ref{onedim} and is given by Eq.~(\ref{nuT}). Number fluctuations at finite temperature are obtained from the integral (\ref{fluct1D}). Numerical integration shows that at finite temperature the additivity is restored and number fluctuations are proportional to $R$.

Notice, in conclusion of this section, that the asymptotic law $\left\langle \delta N^2 \right\rangle \sim R^2 \ln R $ is valid also for an arbitrary non-superfluid Fermi liquid, because one has always $S(k) \sim k $ at $k \to 0$. The prefactor in this equation can be calculated in terms of the $f$-function of the liquid by solving the Landau kinetic equation (see Sec.~91 in \cite{Lifshitz80}). However the coefficient inside the logarithm depends on the high momentum behavior and cannot be calculated in a general form.

\section{Conclusion}
In conclusion we have investigated in this paper the zero-point fluctuations of the particle number in a macroscopically large sphere of radius $R$ for different types of superfluids. In all three-dimensional cases at zero temperature we obtain an anomalous dependence $\left\langle \delta N^2 \right\rangle \sim R^2 \ln R $. The prefactor in this equation depends on the system, but can be calculated in general form, while the coefficient inside the logarithm depends on the high momentum behavior of the system under consideration. We have calculated this last coefficient explicitly throughout the BEC-BCS crossover, and in particular for the limiting case of the weakly interacting Bose gas. We have also considered the somewhat singular limit of the non-interacting Fermi gas. Our obtained results can be relevant for noise measurement experiments in cold gases.

We thank S. Stringari for useful discussions and Yu. M. Tzypeniuk for pointing out Ref.~\cite{Price54}. We thank B. Jackson for reading the manuscript. G.E.A. acknowledges support from MEC (Spain).
\appendix

\section{Derivation for the model structure factor and for the weakly interacting Bose gas}
\label{mod}

From Eqs.~(\ref{dNTF}) and (\ref{model}) we have to calculate, with the change of variable $x=kR$
\begin{eqnarray}
\int\limits_{0}^{\infty} dk\frac{S_0(k)}{k^4} \left[\sin(kR)-kR \cos(kR)\right]^2 = \alpha R^2 I_1 + R^3 I_2,
\end{eqnarray}
where
\begin{eqnarray}
\hspace{1mm} I_1 = \int\limits_{0}^{R/\alpha} dx \frac{1}{x^3} \left[\sin x-x \cos x\right]^2,
\end{eqnarray}
\vspace{-5mm}
\begin{eqnarray}
I_2 = \int\limits_{R/\alpha}^{\infty} dx \frac{1}{x^4} \left[\sin x-x \cos x\right]^2.
\end{eqnarray}
Since we are interested in the limit $R\rightarrow \infty$, it is clear that in $I_2$ only the contribution from $x \cos x$ is left at the dominant order in $R$ we are interested in. Moreover, we can, for the same reason, replace $\cos^2 x$ by its average $1/2$, which gives to dominant order
\begin{eqnarray}
I_2=\frac{1}{2}\frac{\alpha}{R}.
\end{eqnarray}
It is in principle more complicated to handle $I_1$ since the dominant $\ln R$ term comes from the contribution of $x \cos x$ at the upper bound, the other terms becoming negligible in this limit, while for $x \rightarrow 0$ all the terms need to be retained in order to avoid singularities. Nevertheless this can be conveniently done by integrating by parts:
\begin{eqnarray}
\hspace{-30mm}I_1=-\frac{1}{2}\cos^2\frac{R}{\alpha } +  \int\limits_{0}^{R/\alpha } dx\frac{\sin x (\sin x - x \cos x)}{x} \nonumber
\end{eqnarray}
\vspace{-5mm}
\begin{eqnarray}
=-\frac{1}{2}+\frac{1}{2}  \int\limits_{0}^{2R/\alpha } dt \frac{1-\cos t}{t} = \frac{1}{2}[\ln \frac{2R}{\alpha } + \gamma - \Ci (\frac{2R}{\alpha}) -1].
\end{eqnarray}
Here we have omitted in the first step terms going to zero when $R\rightarrow \infty$ since these terms are indeed negligible at the order we are interested in. The last step introducing the cosine integral $\Ci (x)$ can be found in Ref.~\cite{gr}. Since $\Ci (x) \rightarrow 0$ when $x \rightarrow \infty$ we obtain by gathering all the terms
\begin{eqnarray}
\alpha R^2 I_1 + R^3 I_2 = \frac{1}{2} \alpha R^2\; \ln \frac{2 e^\gamma R}{\alpha }.
\end{eqnarray}
This gives for the number fluctuation
\begin{eqnarray}
\left\langle \delta N^2\right\rangle=8{\bar n}\alpha R^2\ln \frac{2 e^\gamma R}{\alpha }=8{\bar n}\alpha R^2 \left[\;\ln \frac{R}{\alpha } + 1.270\;\right],
\end{eqnarray}
as indicated in the text.

To handle the bosonic case where
\begin{eqnarray}
S_B(k) = \frac{\hbar k}{2m_B c}\frac{1} {\sqrt{1+(\frac{\hbar k}{2m_B c})^2}}
\end{eqnarray}
we proceed by the difference with the above model, where we take naturally $\alpha =\hbar /(2m_B c)$, just as indicated in Sec.~\ref{secBCSBEC}. We are left with the calculation of:
\begin{eqnarray}
\hspace{-30mm}\left\langle \delta N^2\right\rangle-\left\langle \delta N^2\right\rangle_{\mathrm {mod}}= 4 {\bar n} R^2 \left[\; \int\limits_{0}^{1/\alpha } dk\frac{S(k)-\alpha k}{k^2} + \int\limits_{1/\alpha }^{\infty} dk\frac{S(k)-1}{k^2} \; \right].
 \end{eqnarray}
With the change of variable $k=2m_B c x/\hbar= x/\alpha $ we are left to evaluate the simple integrals:
\begin{eqnarray}
\int\limits_{0}^{1} dx(\frac{1}{x\sqrt{1+x^2}}-\frac{1}{x})=\ln 2 - \ln (1+\sqrt{2}) \\ \int\limits_{1 }^{\infty} dx(\frac{1}{x\sqrt{1+x^2}}-\frac{1}{x^2})=\ln (1+\sqrt{2})-1,
 \end{eqnarray}
 which leads to the result indicated in the text:
 \begin{eqnarray}
\left\langle \delta N^2\right\rangle=4{\bar n}_B\alpha R^2\ln \frac{4 e^{\gamma-1} R}{\alpha }.
\end{eqnarray}

\section{Derivation in an ideal Fermi gas}\label{idealferm}

The same-spin density-density correlation function $\nu_{\uparrow \uparrow}(r)$ is given by Eq.~(\ref{nu:IFG}), from which the number fluctuations is obtained from Eqs.~(\ref{fluct}) and (\ref{tau}):
\begin{eqnarray}\label{b1}
\left\langle \delta N_{\uparrow \uparrow}^2\right\rangle &
=&4\pi \bar{n}\int\limits_0^{2R}\nu_{\uparrow \uparrow }(r)\tau(r)r^2dr= \nonumber \\
&=&\frac{4\pi \bar{n}R^3}3 +\frac{4}{3\pi ^2}\int\limits_0^2 \frac{dx}{x^4}\left(-1+\frac{3x}{4}-\frac{x^3}{16}\right)[\sin (k_{F}Rx)-(k_{F}Rx) \cos (k_{F}Rx)]^2,
\end{eqnarray}
where the first term comes from the $\delta^3(r)$ contribution in $\nu_{\uparrow \uparrow }(r)$ in Eq.~(\ref{nu:IFG}) and we have just made the change $r=Rx$. Taking into account
\begin{eqnarray}
\int\limits_0^\infty \frac{dx}{x^4}[\sin (k_{F}Rx)-(k_{F}Rx) \cos (k_{F}Rx)]^2 = \frac{\pi }{6} (k_FR)^3,
\end{eqnarray}
leading to a contribution which cancels the first term of Eq.~(\ref{b1}), we find:
\begin{eqnarray}\label{b3}
\left\langle \delta N_{\uparrow \uparrow}^2\right\rangle &
=&\frac{4}{3\pi ^2}\int\limits_0^2 \frac{dx}{x^4}\left(\frac{3x}{4}-\frac{x^3}{16}\right)[\sin (k_{F}Rx)-(k_{F}Rx) \cos (k_{F}Rx)]^2 \nonumber \\
&+&\frac{4}{3\pi ^2}\int\limits_2^\infty \frac{dx}{x^4}[\sin (k_{F}Rx)-(k_{F}Rx) \cos (k_{F}Rx)]^2.
\end{eqnarray}
This expression can also be obtained directly from Eq.~(\ref{dNTF}), together with the expression of the structure factor for the ideal Fermi gas $S(k)=(3k/4k_F)-k^3/(16 k_{F}^{3})$ for $k<2k_F$ and $S(k)=1$ for $k>2k_F$.

Setting $\bar{R}=4k_{F}R$, we obtain the following explicit expression \cite{gr}:
\begin{eqnarray}\label{IFGexact}
\left\langle \delta N_{\uparrow \uparrow}^2\right\rangle = \frac{\bar{R}^2}{32\pi ^2}\left[A(\bar{R})-\frac{1}{2}\right]+
\frac{\bar{R}^2}{144\pi^2}\left[\frac{\pi \bar{R}}2-\bar{R}\Si(\bar{R})-\cos \bar{R} \right] \nonumber \\
+ \frac{1}{288\pi^2}\left[7\bar{R} \sin \bar{R} -12 A(\bar{R}) -5 \cos \bar{R}+5\right],
\end{eqnarray}
where $\Si(x)=\int\limits_0^x\frac{\sin t}{t}\;dt$ and $\Ci(x)=-\int\limits_x^\infty\frac{\cos t}{t}\;dt$ are sine and cosine integral functions and $A(\bar{R})=\ln \bar{R}+\gamma-\Ci(\bar{R})$, with the dominant behavior $\ln \bar{R}+\gamma$ for large $\bar{R}$.

The leading term in a sphere of large radius $R$ is derived from the first term and is indeed
\begin{eqnarray}
\left\langle \delta N_{\uparrow \uparrow}^2\right\rangle =\frac{(k_{F}R)^2}{2\pi^2}\ln \left[4e^{\gamma -\frac{1}2}k_{F}R\right]+\mathcal{O}(k_{F}R).
\end{eqnarray}

\section{Density fluctuations of one-dimensional ideal Fermi gas at a finite
temperature}\label{onedim}

It is of interest to discuss the effect of temperature. The one-dimensional ideal Fermi gas provides a simple example where this can be done. A nice expression for $\nu(r)$ at low temperature can be derived, but still one has to calculate an integral involving $\nu(r)$ in order to obtain $\left\langle \delta N^2\right\rangle$.

The density-density correlation function is written in second quantization as
\begin{eqnarray}
{\bar n}\nu ({\bf x},{\bf x}^{\prime})=\left\langle n({\bf x})n({\bf x}^{\prime})\right\rangle-{\bar n}^2 =\left\langle \Psi^{\dagger}({\bf x}) \Psi({\bf x})\Psi^{\dagger}({\bf x}^{\prime})\Psi({\bf x}^{\prime})\right\rangle-{\bar n}^2,
\end{eqnarray}
or, with $\Psi ({\bf x})=\sum_{{\bf k}}e^{i{\bf k}{\bf x}} a_{{\bf k}}/\sqrt{V}$, we have in terms of plane waves creation and annihilation operators:
\begin{eqnarray}
{\bar n}\nu ({\bf x},{\bf x}^{\prime}) =\frac{1}{V^2}\sum\limits_{{\bf k}_1,{\bf k}_2,{\bf k}_3,{\bf k}_4}\left\langle a_{{\bf k}_1}^{\dagger}a_{{\bf k}_2}a_{{\bf k}_3}^{\dagger}a_{{\bf k}_4} \right\rangle e^{i{\bf x}({\bf k}_2-{\bf k}_1)}e^{i{\bf x}^{\prime}({\bf k}_4-{\bf k}_3)}-{\bar n}^2. \label{nusecondq}
\end{eqnarray}
Using Wick's theorem we get $\left\langle a_{{\bf k}_1}^{\dagger}a_{{\bf k}_2}a_{{\bf k}_3}^{\dagger}a_{{\bf k}_4}\right\rangle =\left\langle a_{{\bf k}_1}^{\dagger}a_{{\bf k}_2}\right\rangle\left\langle a_{{\bf k}_3}^{\dagger}a_{{\bf k}_4}\right\rangle +\left\langle a_{{\bf k}_1}^{\dagger}a_{{\bf k}_4}\right\rangle \left\langle a_{{\bf k}_2}a_{{\bf k}_3}^{\dagger}\right\rangle = n_{{\bf k}_1}n_{{\bf k}_3}\delta_{{\bf k}_1,{\bf k}_2}\delta_{{\bf k}_3,{\bf k}_4} + n_{{\bf k}_1}\delta_{{\bf k}_1,{\bf k}_4}(1-n_{{\bf k}_2}\delta_{{\bf k}_2,{\bf k}_3})$. In a homogeneous system the correlator (\ref{nusecondq}) depends only on the relative distance ${\bf r}={\bf x}-{\bf x}^{\prime}$:
\begin{eqnarray}
{\bar n}\nu ({\bf r}) ={\bar n}\delta^D ({\bf r})-\left\vert\frac{1}{V} \sum\limits_kn_ke^{i{\bf k}{\bf r}}\right\vert^2 ={\bar n}\delta^D ({\bf r})-\left\vert\int n_ke^{i{\bf k}{\bf r}}\frac{d^Dk}{(2\pi)^D} \right\vert^2,
\end{eqnarray}
where $D$ is the space dimension. For free fermions $n_k=\left[e^{(\varepsilon_k-\mu )/T}+1\right]^{-1}$. At low temperature the chemical potential can be well-approximated by the zero temperature value $\mu =\frac{\hbar^2k_F^2}{2m}$. In a one-dimensional system one has to evaluate the following integral:
\begin{eqnarray}
I=2\int\limits_0^\infty\frac{\cos kr}{e^{\frac{\hbar^2(k^2-k_F^2)}{2mT}}+1}\frac{dk}{2\pi}. \label{int1D}
\end{eqnarray}

Let us first consider the $T=0$ case. We find easily, as indicated in the text
\begin{eqnarray}\label{nutzero}
\nu(r)=\delta(r)-\bar{n} \frac{\sin^2(k_F r)}{(k_F r)^2}.
\end{eqnarray}
The strength of fluctuation can then be obtained exactly by calculations analogous to the ones performed in Appendix~\ref{idealferm}. We find, with the same notation $\bar{R}=4k_FR$ as in Appendix~\ref{idealferm}:
\begin{eqnarray}
\pi ^2 \left\langle \delta N^2\right\rangle = \ln\bar{R} +\gamma+1 +\frac{\pi\bar{R}}{2} -\bar{R}\Si(\bar{R}) -\Ci(\bar{R}) -\cos(\bar{R}), \label{dN2_1D_T0}
\end{eqnarray}
where $\Ci(x)$ and $\Si(x)$ are cosine and sine integral functions respectively. The first three terms in Eq.~(\ref{dN2_1D_T0}) provide the dominant contribution and are reported in Sec.~\ref{secIFG}. The contribution of remaining terms scales like $-\cos(\bar{R})/\bar{R^2}$ for $\bar{R}\gg 1$. Indeed, there are canceling contributions as can be seen from asymptotic expansion $\Si(\bar{R})=\pi/2-\cos(\bar{R})/\bar{R}-\sin(\bar{R})/\bar{R}^2 +2\cos(\bar{R})/\bar{R}^3+{\cal O}(\bar{R}^{-4})$, and $\Ci(\bar{R})= \sin(\bar{R})/\bar{R}-\cos(\bar{R})/\bar{R}^2+{\cal O}(\bar{R}^{-3})$.

We turn now to the evaluation of $\nu(r)$ at low temperature. As can be seen from integration by parts, the main contribution to the integral (\ref{int1D}) comes from $k \approx k_F$. We proceed with the integration in a dimensionless variable $x=\hbar^2(k^2-k_F^2)/2mT\approx\hbar^2k_F(k-k_F)/mT$ and extend the lower limit of integration in Eq.~(\ref{int1D}) to minus infinity, thus adding exponentially small contributions.

The resulting integral can be integrated by parts,
\begin{eqnarray}
I = \frac{k_\lambda}{\pi}\int\limits_{-\infty}^{\infty}\frac{\cos(k_Fr+k_\lambda rx)}{e^x+1}dx = \frac{1}{\pi r}\int\limits_{-\infty}^{\infty}\frac{\sin (k_Fr+k_\lambda rx)}{(e^x+1)(e^{-x}+1)}dx = \frac{k_\lambda\sin k_F r}{\sinh\pi k_\lambda r},
\end{eqnarray}
where $k_\lambda =\frac{mT}{\hbar^2k_F}$ is a characteristic thermal momentum of the problem. Now we can write the density-density correlator:
\begin{eqnarray}
\nu(r)=\delta (r)-{\bar n} \left(\frac{\pi k_\lambda\sin k_Fr}{k_F\sinh\pi k_\lambda r}\right)^2. \label{nuT}
\end{eqnarray}
This shows that the density-density correlation function in a one-dimensional Fermi gas at finite temperature has an exponential decay at large $r$ (like in the 3D case at finite temperature), in contrast with the $T=0$ result, Eq.~(\ref{nutzero}), exhibiting a power law decay recovered from Eq.~(\ref{nuT}) in the limit of zero temperature $k_\lambda \rightarrow 0$.


\begin{thebibliography}{7}
\bibitem[{\citenamefont{}()}]{rem}

\bibinfo{author} The periodic function appearing in these oscillations is essentially the same as the
one in Eq.(\ref{TFnu}), giving the expression of $T(k)$. This is due to the fact that the $S(k)$ is essentially the overlap of two Fermi spheres in $k$-space \cite{Pines89} and $\nu_{\uparrow\uparrow}(r)$ is its Fourier transform.
\end{thebibliography}
\end{document}